\documentclass[twocolumn,superscriptaddress,amsmath,ammsymb]{revtex4-1}
\usepackage{epsfig}
\usepackage{amsmath}
\usepackage{amssymb}
\usepackage{graphicx}
\usepackage{dcolumn}
\usepackage{bm}
\usepackage[usenames]{color}
\usepackage[breaklinks]{hyperref}
\usepackage{soul}
\usepackage{ulem}
\hypersetup{colorlinks=true,allcolors=blue}

\begin{document}
	
\title{Quantum oscillations and three-dimensional quantum Hall effect in ZrTe$_5$}

\author{Yi-Xiang Wang}
\email{wangyixiang@jiangnan.edu.cn}
\affiliation{School of Science, Jiangnan University, Wuxi 214122, China}
\affiliation{School of Physics and Electronics, Hunan University, Changsha 410082, China}

\author{Zhigang Cai}
\affiliation{School of Science, Jiangnan University, Wuxi 214122, China}
 
\date{\today}

\begin{abstract}
Recent experiments have reported a lot of spectacular transport properties in topological materials, such as quantum oscillations and three-dimensional (3D) quantum Hall effect (QHE) in ZrTe$_5$. 
In this paper, by using a strong topological insulator model to describe ZrTe$_5$, we study the magnetotransport property of the 3D system. 
With fixed carrier density, we find that there exists a deferring effect in the chemical potential, which favors distinguishing the saddle points of the inverted LLs.  
On the other hand, with fixed chemical potential, the features of 3D QHE are demonstrated and we attribute the underlying mechanisms to the interplay between Dirac fermions, magnetic field and impurity scatterings.  
\end{abstract} 

\maketitle

\section{Introduction}

In the past two decades, the search of various topological phases in materials has ignited ongoing interest in condensed matter physics~\cite{M.Z.Hasan, X.L.Qi}.  Compared to conventional trivial phases, the most exotic property of topological phases is that they can support the surface states, which show robustness to the weak disorder or defects.  To detect the topological phases in experiments, one may capture the surface states by tuning the chemical potential inside the gap~\cite{B.Q.Lv, J.A.Sobota}.  This works well for wide-gap topological insulators (TIs), such as Bi$_2$Se$_3$ and Bi$_2$Te$_3$, with a gap larger than $200$ meV~\cite{Y.Xia, H.Zhang, Y.L.Chen, C.X.Liu}.  But for a narrow-gap material, it is not easy to tune the chemical potential inside the gap.  

Three-dimensional (3D) ZrTe$_5$ is such a narrow-gap topological material~\cite{H.Weng}.  Recent experiments show that it can exhibit a variety of unique transport properties, including chiral magnetic effect~\cite{Q.Li}, anomalous Hall effect~\cite{T.Liang2018}, saturating thermoelectric Hall effect~\cite{J.L.Zhang, W.Zhang}, and gigantic magnetochiral anisotropy~\cite{Y.Wang}.  
Instead of the surface states, an alternative route to probe the topological phase in ZrTe$_5$ was proposed through detecting the saddle points of the inverted bands~\cite{J.Wang}.  Under a magnetic field, electrons are forced to move on quantized curved orbits-the Landau levels (LLs), which, if inverted, can exhibit two saddle points in the Brillouin zone (BZ)~\cite{Y.Jiang, L.You}, the $\Gamma$ point and the $\zeta$ point.  When the magnetic field changes, the chemical potential will cross the LLs one by one, leading to quantum oscillations (QOs) in transport and thermodynamics~\cite{D.Shoeberg, G.M.Monterio, S.Kaushik}, in which the signatures of the saddle points need further study. 

3D quantum Hall effect (QHE) was another intriguing phenomenon observed recently in ZrTe$_5$~\cite{F.Tang, S.Galeski2020} and HfTe$_5$~\cite{P.Wang, S.Galeski2020}.  The Hall resistivity $\rho_{xy}$ was reported to exhibit a quasi-quantized plateau in the quantum limit.  The quantum limit means that only the zeroth LL is occupied, which can be reached at a critical magnetic field of about 1.3 T in ZrTe$_5$~\cite{F.Tang} and 1.8 T in HfTe$_5$~\cite{P.Wang}, both indicating the low carrier densities in these 3D crystals.   More importantly, the plateau scales as  $\rho_{xy}\simeq\frac{\pi}{k_F}\frac{h}{e^2}$, with $k_F$ being the Fermi wavevector.  It was proposed that the enhanced interactions in the quantum limit can induce the Fermi surface instability, which in turn drives the charge density wave (CDW) that would open a gap around the Fermi level~\cite{F.Tang, P.Wang}.  In a 3D system, applying a magnetic field can reduce the energy spectrum to one dimension (1D).  Since the 1D system features an almost perfect nesting Fermi surface, its ground state favors the CDW states~\cite{G.Gruner}, thus it seems that the CDW picture may be supported in explaining 3D QHE~\cite{F.Qin}. 

Nevertheless, there exist several facts against such a CDW scenario: (i) if the Fermi energy lies in the gap, the quasi-quantized Hall plateaus should correspond to the vanishing longitudinal resistivity $\rho_{xx}$, which, however, remains finite as observed in ZrTe$_5$~\cite{F.Tang, S.Galeski2020} and HfTe$_5$~\cite{P.Wang, S.Galeski2020};  (ii) the CDW picture can only account for the plateau in the quantum limit, but not for the other plateaus in the QO  regime~\cite{F.Tang, S.Galeski2020, P.Wang};  and (iii) in a recent study~\cite{S.Galeski2021}, Galeski \textit{et al}., corroborated that the CDW states were absent in the ZrTe$_5$ crystal through various measurements, based on which they suggested that 3D QHE emerges from the interplay between the intrinsic properties of ZrTe$_5$ and its Dirac-type semimetal character.  Thus, the physical mechanism behind 3D QHE also requires further study.  

Motivated by this progress, in this paper, we use the strong TI model to describe the low-energy excitations in ZrTe$_5$ and systematically investigate its bulk magneotransport, with a focus on the problems of QOs and 3D QHE.  
We will consider both conditions of fixed carrier density and fixed chemical potential.  The former condition was assumed in understanding the saturating thermoelectric effect~\cite{J.L.Zhang}, anomalous resistivity at finite temperature~\cite{B.Fu, C.Wang} in ZrTe$_5$, whereas the latter one was used to explain the optical spectroscopy~\cite{B.Xu, E.Martino} and magnetoinfared spectroscopy measurements~\cite{R.Y.Chen, Y.Jiang} in ZrTe$_5$. 
We will calculate the density of states (DOS), the oscillating chemical potential or carrier density, as well as the conductivities and resistivities in the 3D system.  
In the calculations, the impurity scatterings are included in the system phenomenologically by introducing the linewidth broadening to the LLs.  

Our results show that under the condition of fixed carrier density, there exists a deferring effect in the chemical potential, which favors distinguishing the saddle points of the inverted LLs.  At weak impurity scatterings, the anomalous peak in the quantum limit and the fourfold peaks in the QO regime can be captured in the DOS as well as the longitudinal resistivity $\rho_{xx}$, which can help identify the band inversions of the LLs.  
Under the condition of fixed chemical potential, the characteristics of 3D QHE are demonstrated, including the quasi-quantized plateau in the Hall resistivity $\rho_{xy}$ and the nonvanishing dips in $\rho_{xx}$.  We attribute 3D QHE to the interplay between Dirac fermions, magnetic field, and impurity scatterings. 
Our paper could provide more insights to understand the peculiar transport properties of ZrTe$_5$ in experiments.

\section{Model} 

In the study of ZrTe$_5$, a simple two-band model was proposed to calculate the experimental observables~\cite{Y.X.Wang2020a, Z.Rukelj}, and also a four-band model based on the \textit{ab initio} approach was used to investigate the optical conductivity in the system~\cite{C.Morice}.  
Since it is generally believed that the ground state of 3D ZrTe$_5$ lies in the proximity of the phase boundary between a strong TI and a weak TI~\cite{H.Weng}, here we adopt the strong TI model to describe its low-energy excitations.  The strong TI phase 
features the in-plane and out-of-plane band inversions, and has been supported by many experimental evidences in ZrTe$_5$~\cite{G.Manzoni, B.Xu, Y.Jiang, J.Wang, Z.G.Chen}. 

In the four-component basis $(|+\uparrow\rangle, |-\uparrow\rangle, |+\downarrow\rangle, |-\downarrow\rangle)^T$, the Hamiltonian reads ($\hbar=1$)~\cite{Y.Jiang, J.Wang, H.Zhang, C.X.Liu}
\begin{align}
H(\boldsymbol k)
=&v(k_x\sigma_z\otimes\tau_x+k_yI\otimes \tau_y)
+v_z k_z\sigma_x\otimes\tau_x
\nonumber\\
&+[M-\xi(k_x^2+k_y^2)-\xi_zk_z^2]I\otimes \tau_z,
\label{H0}
\end{align} 
where $\sigma$ and $\tau$ are the Pauli matrices acting on the spin and orbit degrees of freedom, respectively.  $v$ and $v_z$ are the Fermi velocities, $\xi$ and $\xi_z$ are the band inversion parameters, and $M$ denotes the Dirac mass.  In the following calculations, we take the parameters from the magnetoinfrared measurements~\cite{Y.Jiang}: $(v,v_z)=(6,0.5)\times10^5$ m/s, $(\xi,\xi_z)=(100,200)$ meV nm$^2$, $M=7.5$ meV.  With these parameters, besides the $\Gamma$ point, $H(\boldsymbol k)$ owns the second saddle point$-\zeta$ point, which is located at  $(k_\parallel,k_z)=\big(0,(\frac{M}{\xi_z}-\frac{v_z^2}{2\xi_z^2})^\frac{1}{2}\big)$, with  $k_\parallel=\sqrt{k_x^2+k_y^2}$.  In addition, $H(\boldsymbol k)$ can support more topological phases when the parameters are taken in a broad range, such as single Dirac point phase, Dirac ring phase~\cite{Y.X.Wang2021}, weak TI phase~\cite{W.Wu} and Dirac semimetal phase. 

Under a perpendicular magnetic field $\boldsymbol B=B\boldsymbol e_z$, the quantized LLs are formed, with dispersions along the $z$ direction.  The magnetic field is included in the system with the help of the Peierls substitution, $\boldsymbol k\rightarrow \boldsymbol\pi-e\boldsymbol A$, where the vector potential is chosen as  $\boldsymbol A=(0,-By,0)$ in the Landau gauge.  We then introduce the ladder operators  $a^\dagger=\frac{l_B}{\sqrt2}(\pi_x-i\pi_y)$ and $a=\frac{l_B}{\sqrt2}(\pi_x+i\pi_y)$, with the magnetic length $l_B=\frac{1}{\sqrt{eB}}=\frac{25.6}{\sqrt B}$ nm.  By using the trial wavefunction $\psi_n=(c_n^1|n\rangle, c_n^2|n-1\rangle, c_n^3|n-1\rangle, c_n^4|n\rangle)^T$, in which the state $|n\rangle$ is defined as $a^\dagger a|n\rangle=n|n\rangle$ and $c_n^{1,\cdots,4}$ denote the coefficients, the energies for the zeroth and $n\geq1$ LLs are obtained as~\cite{L.You} 
\begin{align} 
\varepsilon_{0s}(k_z)=&s\Big[\Big(M-\frac{\xi}{l_B^2}-\xi_zk_z^2\Big)^2
+v_z^2k_z^2\Big]^{\frac{1}{2}}, 
\\
\varepsilon_{ns\lambda}(k_z)=&s\Big[\Big(\sqrt{\Big(M-\frac{2 n\xi}{l_B^2} -\xi_zk_z^2\Big)^2+\frac{2nv^2}{l_B^2}}+\frac{\lambda\xi}{l_B^2}\Big)^2
\nonumber\\
&+v_z^2 k_z^2\Big]^\frac{1}{2}, 
\label{epsilonn} 
\end{align}
respectively, where the index $s=\pm1$ denotes the conduction/valence band, and $\lambda=\pm1$ the two branches representing the Zeeman splitting between up spin and down spin~\cite{L.You}.  With increasing magnetic field, we see that the zeroth LL moves to zero energy, while the $n\geq1$ LLs move away from zero energy.  The latter is because in Eq.~(\ref{epsilonn}), the second term $\frac{2nv^2}{l_B^2}$ under the square root is much larger than the first squared term. 

The band inversions can also exist in LLs.  The inverted $\zeta$ points are located at 
\begin{align}
k_{z0}=\Big(\frac{M}{\xi_z}-\frac{\xi}{\xi_z l_B^2}
-\frac{v_z^2}{2\xi_z^2}\Big)^\frac{1}{2}, 
\label{kz0}
\\
k_{zn}\simeq\Big(\frac{M}{\xi_z}-\frac{2n\xi}{\xi_z l_B^2}-\frac{v_z^2}{2\xi_z^2}\Big)^\frac{1}{2}, 
\label{kzn} 
\end{align} 
for the zeroth and $n\geq1$ LL, respectively.  Note that in $k_{zn}$, the weak $\lambda$ term has been  neglected.  

At a strong magnetic field, the inverted LLs may become noninverted~\cite{L.You}, implying that the $\zeta$ point would be merged with the $\Gamma$ point.  For the zeroth LL, the critical magnetic field is $B_0^c=\frac{2M\xi_z-v_z^2}{2e\xi\xi_z}=31.5$ T, and for the $n\geq1$ LL, $B_n^c=\frac{2M\xi_z-v_z^2}{4ne\xi\xi_z}=\frac{15.8}{n}$ T.  Below we see that the chosen magnetic fields are much weaker than these critical values and thus the LLs are always inverted.

\section{Method}

\subsection{Density of states and carrier density}

We calculate the DOS of 3D ZrTe$_5$ under a magnetic field.  With the help of the retarded Green's function $G^R(\varepsilon)=(\varepsilon-H+i\eta)^{-1}$~\cite{Y.X.Wang2021}, the DOS is defined as
\begin{align}
D(\varepsilon)&=-\frac{g}{\pi L_z}\sum_{k_z} 
\text{Tr}[\text{Im}G^R(\varepsilon,k_z)]
\nonumber\\
&=\frac{g}{\pi L_z}\sum_{k_z}\sum_{ns\lambda} 
\frac{\eta}{[\varepsilon-\varepsilon_{ns\lambda}(k_z)]^2+\eta^2}, 
\end{align} 
where $L_z$ is system size in the $z$ direction, $g=\frac{1}{2\pi l_B^2}$ is the LL degeneracy in the $x$-$y$ plane and can be denoted as the uniform DOS, and $\eta$ represents the linewidth broadening phenomenologically that is induced by impurity scatterings.  Consider the short-range pointlike impurities, $U=u_0\sum_i \delta(\boldsymbol r-\boldsymbol R_i)$, and the impurity concentration $n_i$, then in the Born approximation, the scattering time $\tau=\frac{1}{2\eta}=\frac{1}{2\pi\gamma D_F}$~\cite{P.Hosur, A.A.Burkov2014, J.Klier,  H.W.Wang}, with $\gamma=\frac{1}{2}n_i u_0^2$ characterizing the strength of the impurity potential, and $D_F$ the DOS at the Fermi energy.  Note that when the band inversions are absent, $\xi=\xi_z=0$, the divergence in the DOS is one over the square-root type~\cite{S.Kaushik, J.Klier, Y.X.Wang2019}.  Now the LL energies show complicated dependencies on $k_z$, leading to the intricated behavior of the DOS, as shown below. 

On the other hand, the carrier density $n_0$ is related to the DOS by~\cite{Ashby}   
\begin{align}
n_0(\mu,B)=\int_0^\infty d\varepsilon D(\varepsilon)f(\varepsilon)
+\int_{-\infty}^0 d\varepsilon D(\varepsilon)[1-f(\varepsilon)],  
\label{n0} 
\end{align}
where $f(\varepsilon)=\big[\text{exp}(\frac{\varepsilon-\mu}{k_BT})+1\big]^{-1}$ is the Fermi distribution function, with $\mu$ being the chemical potential, $k_B$ the Boltzmann constant, and $T$ the temperature.  In the following, we focus on zero temperature.

\subsection{Conductivity and resistivity}

We also calculate the conductivity and resistivity of the system.  The conductivity tensors $\sigma_{\alpha\beta}$ can be derived from the linear-response Kubo-Streda formula~\cite{L.Smrcka, G.D.Mahan}, 
\begin{align}
\sigma_{\alpha\beta}=&\frac{1}{2\pi V} \sum_{\boldsymbol k}
\int_{-\infty}^\infty d\varepsilon f(\varepsilon)\Big[
\text{Tr}\big(J_\alpha\frac{dG^R}{d\varepsilon}J_\beta (G^A-G^R)
\nonumber\\
&-J_\alpha (G^A-G^R) J_\beta \frac{dG^A}{d\varepsilon}
\big)\Big],    
\label{Kubo-Streda}
\end{align}
where $V$ is the system volume, $J_\alpha=e\frac{\partial H}{\partial k_\alpha}$ is the current density operator, and $G^A(\varepsilon)=(\varepsilon-H-i\eta)^{-1}$ is the advanced Green's function.  In the transport theory~\cite{S.Datta}, the relaxation time denotes the time that an electron experiences from the undisturbed state to the equilibrium state through impurity scatterings, while the inverse LL broadening is related to the lifetime that an electron stays in the state.  Here we assume that both quantities are equal to the scattering time $\tau$~\cite{S.Galeski2021}.  Such an assumption has the advantage that the conductivities can be directly determined from the band structures.  

After a straightforward calculation, we obtain the longitudinal conductivity $\sigma_{xx}$ and transverse Hall conductivity $\sigma_{xy}$, which are expressed as a summation over the LL index (see Appendix A), 
\begin{align}
\sigma_{xx}=&\sigma_0\frac{4g\eta^2}{L_z} 
\sum'\frac{M_{ns\lambda;n+1,s'\lambda'}^2}
{[(\mu-\varepsilon_{ns\lambda})^2+\eta^2][(\mu-\varepsilon_{n+1,s'\lambda'})^2+\eta^2]}, 
\label{sigmaxx}
\end{align}
and
\begin{align}
\sigma_{xy}=&\sigma_0\frac{4g\pi}{L_z} \sum'
\frac{(\varepsilon_{ns\lambda}-\varepsilon_{n+1,s'\lambda'})^2-\eta^2}
{[(\varepsilon_{ns\lambda}-\varepsilon_{n+1,s'\lambda'})^2+\eta^2]^2}
M_{ns\lambda;n+1,s'\lambda'}^2
\nonumber\\
&\times
[\theta(\mu-\varepsilon_{ns\lambda}) 
\theta(\varepsilon_{n+1,s'\lambda'}-\mu)
\nonumber\\
&-\theta(\mu-\varepsilon_{n+1,s'\lambda'}) 
\theta(\varepsilon_{ns\lambda}-\mu)].  
\label{sigmaxy}
\end{align}
where $\sigma_0=\frac{e^2}{h}$ is the unit of the quantum conductivity, $\theta(x)$ is the step function, the summation sign is 
\begin{align}
\sum'=\sum_{k_z}\sum_{n\geq0,s\lambda} \sum_{s'\lambda'},  
\end{align} 
and the matrix element $M_{ns\lambda;n+1,s'\lambda'}$ is 
\begin{align}
&M_{ns\lambda;n+1,s'\lambda'}= 
v(c_{ns\lambda}^1 c_{n+1,s'\lambda'}^2
-c_{ns\lambda}^4 c_{n+1,s'\lambda'}^3)
\nonumber
\\
&+\frac{\sqrt {2(n+1)}\xi}{l_B} (-c_{ns\lambda}^1c_{n+1,s'\lambda'}^1+c_{ns\lambda}^4c_{n+1,s'\lambda'}^4) 
\nonumber\\
&+\frac{\sqrt {2n}\xi}{l_B} (c_{ns\lambda}^2c_{n+1,s'\lambda'}^2-c_{ns\lambda}^3c_{n+1,s'\lambda'}^3). 
\label{matrixelement}
\end{align}
Note that the nonvanishing matrix element $\langle ns\lambda|J_{x/y}|n's'\lambda'\rangle$ determines the common  selection rules $n\rightarrow n\pm1$~\cite{Y.Jiang, L.You}.  

In the transport experiments, since the measured quantities are the resistivities $\rho_{\alpha\beta}$, we need to convert conductivities into resistivities.  Using the relation ${\boldsymbol \sigma}\cdot{\boldsymbol \rho}=I$, we obtain 
\begin{align}
\rho_{xx}=\frac{\sigma_{xx}}{\sigma_{xx}^2+\sigma_{xy}^2}, \quad
\rho_{xy}=\frac{\sigma_{xy}}{\sigma_{xx}^2+\sigma_{xy}^2}.  
\label{rho}
\end{align}

\section{Main Results and Discussions}

\subsection{Fixed carrier density}

In this section, we study the magnetotransport in ZrTe$_5$ under the condition of fixed low carrier density, which is set as $n_0=5.2\times10^{16}$ cm$^{-3}$.  

\begin{figure*} 
	\centering	
	\includegraphics[width=17.2cm]{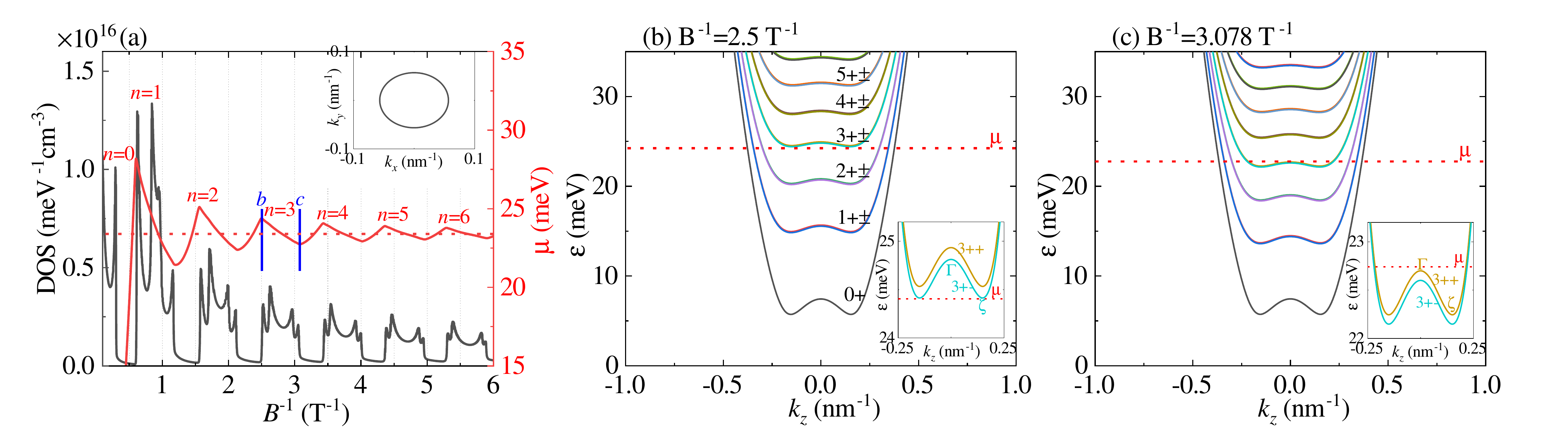}
	\caption{(Color online) (a) The DOS and chemical potential $\mu$ versus the inverse magnetic field $B^{-1}$ with the fixed carrier density $n_0=5.2\times10^{16}$ cm$^{-3}$.  When the magnetic field is absent, the red dotted line indicates the position of the chemical potential $\mu$, and the inset shows the extremal Fermi surface shape in the $k_x$-$k_y$ plane.  (b), (c) The LL dispersions versus $k_z$, with the magnitude of $B^{-1}$ corresponding to the blue lines $b$ and $c$ labeled in (a), respectively.  The insets are enlarged plots of the dispersions around $\mu$.  We set $\eta=0.01$ meV. }
	\label{Fig1}	
\end{figure*} 

The calculated DOS and chemical potential are plotted in Fig.~\ref{Fig1}(a).  We see that when $B^{-1}<0.58$ T$^{-1}$, the system lies in the quantum limit, with an anomalous peak in the DOS.  When $B^{-1}>0.58$ T$^{-1}$, the system lies in the QO regime.  In this regime, the DOS exhibits the fourfold peak structure and the chemical potential oscillates with descending amplitudes.  They both show a period $\Delta_{1/B}=0.95$ T$^{-1}$.  When $B^{-1}$ is asymptotically large (not shown here), the system will enter the semiclassical regime~\cite{H.W.Wang}, in which the energy spacing between two adjacent LLs becomes smaller than the linewidth broadening $\eta$.  

The oscillation period can be understood with the Onsager's relation~\cite{L.Onsager, D.Shoeberg}.  It tells us that the period is inversely proportional to the extremal cross-sectional Fermi surface area $S_e$ in the plane perpendicular to the magnetic field:  
\begin{align}
\Delta_{1/B}^O=\frac{2\pi e}{S_e}.   
\label{period}
\end{align}
When the magnetic field is absent, the chemical potential for the fixed $n_0$ is $\mu_0=23.41$ meV (see Appendix B), as indicated by the red dotted line in Fig.~\ref{Fig1}(a).  The corresponding extremal Fermi surface is plotted in the inset of Fig.~\ref{Fig1}(a).  The Fermi surface is almost an ellipse with the area $S_e=0.01005$ nm$^{-2}$.  Then we obtain $\Delta_{1/B}^O=0.954$ T$^{-1}$, agreeing well with the period extracted from Fig.~\ref{Fig1}(a).  

To illustrate the chemical potential oscillations, we show how $\mu$ crosses the $n=3$ LLs.  The LL dispersions are plotted in Figs.~\ref{Fig1}(b) and~\ref{Fig1}(c), with the magnitudes of $B^{-1}$  corresponding to the blue lines $b$ and $c$ in Fig.~\ref{Fig1}(a), respectively.  Actually with increasing $B^{-1}$, $\mu$ is determined by the competition between the local DOS gaining due to the 1D LL dispersion and the uniform DOS dropping induced by the magnetic field (i.e., the LL degeneracy $g$).  
In Fig.~\ref{Fig1}(b), when $B^{-1}=2.5$ T$^{-1}$, we have $\mu=24.39$ meV, which meets the $\zeta$ point of the $(3+-)$ LL [see Fig.~\ref{Fig1}(b), inset].  
Since the local DOS gaining surpasses the uniform DOS dropping, to conserve the carrier density, $\mu$ will decrease and thus $\mu=24.39$ meV behaves as a peak.  
In Fig.~\ref{Fig1}(c), when $B^{-1}=3.078$ T$^{-1}$, the chemical potential decreases to $\mu=22.74$ meV, which lies above the $\Gamma$ point of the $(3++)$ LL [see Fig.~\ref{Fig1}(c), inset].  With further increasing $B^{-1}$, the local DOS gaining is inferior to the uniform DOS dropping and thus $\mu=22.74$ meV behaves as a valley. 
After that, $\mu$ goes into the next oscillation.  
Importantly, in the local region that includes four saddle points of the $n$-th LLs, the decreasing chemical potential with $B^{-1}$ can defer its crossing over these saddle points.  Such a deferring effect favors distinguishing the saddle points with weak energy difference. 

In the DOS, since the peaks are related to the saddle points of the bands, here the anomalous peak in the quantum limit is caused by the $\Gamma$ point of the zeroth LL, and the fourfold peaks in the $n\geq1$ LLs are attributed to the fact that the $\lambda=\pm1$ branches are splitted and each branch owns two saddle points, the $\Gamma$ point and the $\zeta$ point. 

\begin{figure*} 
	\centering
	\includegraphics[width=12cm]{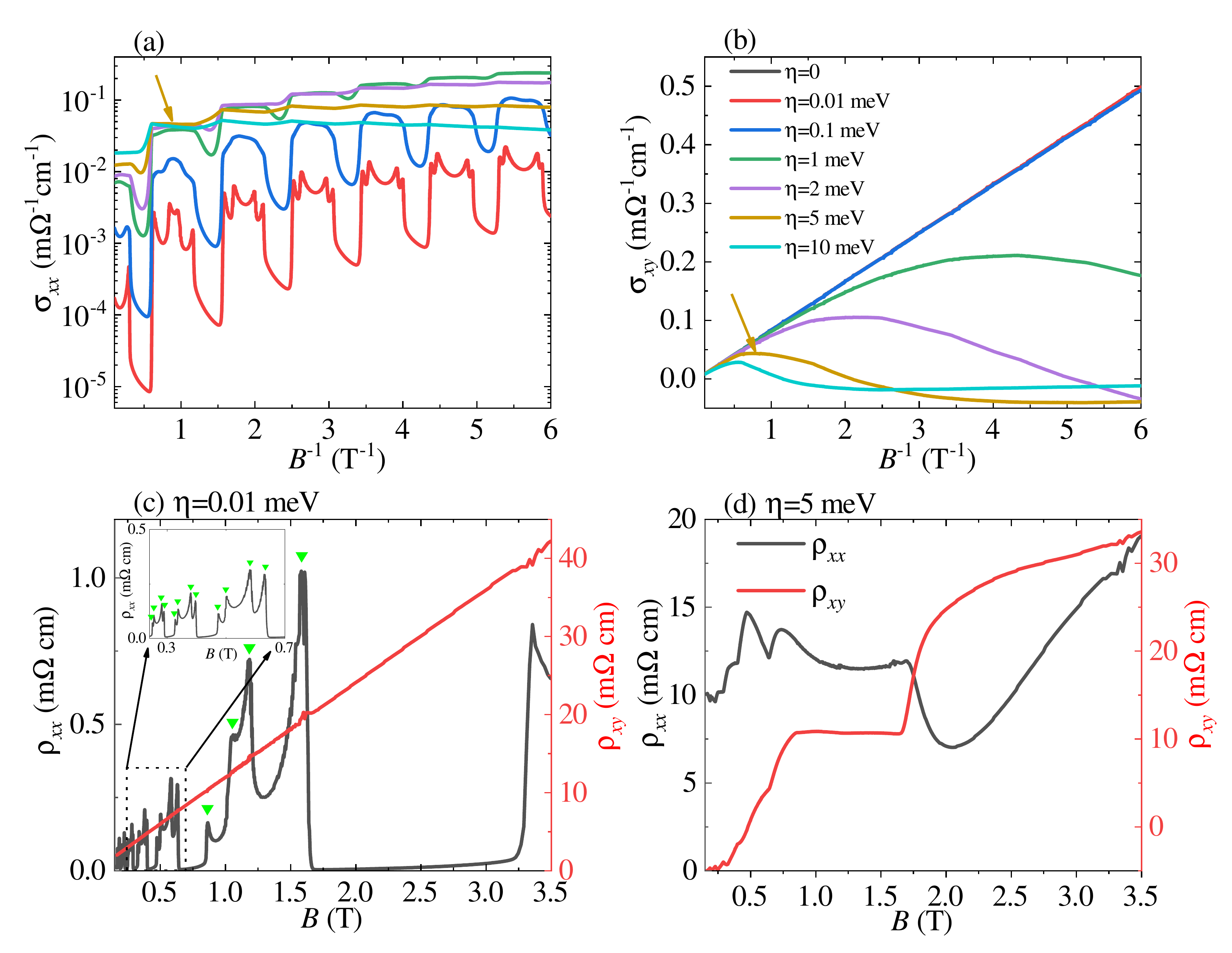}
	\caption{(Color online) The longitudinal conductivity $\sigma_{xx}$ (a) and Hall conductivity $\sigma_{xy}$ (b) versus the inverse magnetic field $B^{-1}$ for different linewidth $\eta$.  In (b), the slope of the line when $\eta=0$ is extracted as $k=0.0831$ m$\Omega^{-1}$cm$^{-1}$T.  The longitudinal resistivity $\rho_{xx}$ and Hall resistivity $\rho_{xy}$ vs the magnetic field $B$ in (c) and (d), with the linewidth $\eta=0.01$ meV and $\eta=5$ meV, respectively.  In (c), the green triangles indicate the fourfold peak structure and the inset shows the enlarged plot framed by the dotted line.  The legends are the same in (a) and (b), as well as in (c) and (d).  The fixed carrier density is $n_0=5.2\times10^{16}$ cm$^{-3}$. } 
	\label{Fig2}	
\end{figure*}

We discuss the saddle point sequence that the chemical potential sweeps with increasing $B^{-1}$.  If 
\begin{align}
\varepsilon_{n+-}(\Gamma)>\varepsilon_{n++}(\zeta), 
\end{align}
the saddle point is swept in the sequence $\zeta$-$\zeta$-$\Gamma$-$\Gamma$.  
By contrast, if
\begin{align}
\varepsilon_{n+-}(\Gamma)<\varepsilon_{n++}(\zeta), 
\end{align}
the sequence will be $\zeta$-$\Gamma$-$\zeta$-$\Gamma$. 
For example, in the inset of Fig.~\ref{Fig1}(b), we have $\varepsilon_{3+-}(\Gamma)>\varepsilon_{3++}(\zeta)$, thus the saddle points are swept in the former sequence, which is also valid for other LLs.  In a previous work~\cite{J.Wang}, the carrier density of the ZrTe$_5$ crystal is reported to be above $10^{17}$ cm$^{-3}$ and thus the quantum limit is reached at the critical magnetic field $B_c>10$ T.  Such a strong magnetic field results in the well-separated $\lambda=\pm1$ branches and the latter sequence $\zeta$-$\Gamma$-$\zeta$-$\Gamma$.  

If we include the spin Zeeman effect into the system, the Hamiltonian is
\begin{align}
H_Z=-\frac{1}{2}g\mu_B B\sigma_z, 
\end{align}
where $g$ denotes the Land\'e $g$-factor and $\mu_B$ is the Bohr magneton.  Taking the value $g\simeq10$ that is reported in ZrTe$_5$~\cite{R.Y.Chen, Z.G.Chen}, the total Zeeman splittings will get strengthened, leading to the further separation of the two $\lambda=\pm1$ branches.  Thus, for the lower LLs, the saddle point sequence may become $\zeta$-$\Gamma$-$\zeta$-$\Gamma$, and for the higher LLs, the sequence remains unchanged.  This means that with increasing LL index $n$, there exists a transition of the saddle point sequence, which depends heavily on the carrier density and the $g$-factor of the LLs. 

Next we study the conductivities.  
The results are plotted as a function of the inverse magnetic field $B^{-1}$ in Figs.~\ref{Fig2}(a) and~\ref{Fig2}(b).  
We analyze the longitudinal conductivity $\sigma_{xx}$ in Fig.~\ref{Fig2}(a).  
In the limiting clean case $\eta=0$, which corresponds to the infinite scattering time, $\sigma_{xx}$ vanishes.  
The impurity-induced scatterings can give rise to the drift current along the electric field direction, leading to the increasing of $\sigma_{xx}$ with weak $\eta$.    
When $\eta=0.01$ meV, $\sigma_{xx}$ exhibits clear oscillations and fourfold peaks that are similar to the DOS. 
When $\eta=0.1$ meV, the fourfold peaks are smoothened and become broad.  
Further increasing $\eta$ to be as strong as $\eta\geq5$ meV, we observe that the oscillations in $\sigma_{xx}$ are smeared out, implying that the LL structures are broken.  
Moreover, in the quantum limit, $\sigma_{xx}$ will get further increased, while in the QO
regime, $\sigma_{xx}$ decreases.  
Such an effect of impurity scatterings on $\sigma_{xx}$ is consistent with our  self-consistent Born approximation study of the diagonal disorder in a Weyl semimetal system~\cite{Y.X.Wang2020}.  

We also analyze the Hall conductivity $\sigma_{xy}$ in Fig.~\ref{Fig2}(b).  When $\eta=0$, $\sigma_{xy}$ is proportional to the inverse magnetic field $B^{-1}$, which retrieves the classical Hall conductivity expression~\cite{V.Konye},
\begin{align}
\sigma_{xy}=\frac{n_0e}{B}. 
\end{align}  
With the extracted slope $k=8.31\times10^{-2}$ m$\Omega^{-1}$ cm$^{-1}$ T, the carrier density is obtained as $n_0=\frac{k}{e}=5.19\times10^{16}$ cm$^{-3}$, which is consistent with the chosen carrier density. 
For the weak impurity scattering $\eta<0.1$ meV, $\eta$ can be ignored in Eq.~(\ref{sigmaxy}), thus $\sigma_{xy}$ shows certain robustness to weak $\eta$.  When $\eta$ increases, the robustness disappears and $\sigma_{xy}$ will get suppressed from the higher LLs to lower ones~\cite{Y.X.Wang2019}.  At strong $\eta\geq5$ meV, $\sigma_{xy}$ even becomes negative at large $B^{-1}$.  This is because when the chemical potential lies between the $n$-th and $(n+1)$-th LLs, the dominate contributions to $\sigma_{xy}$ come from the LL transition $(n,1,\lambda)\rightarrow(n+1,1,\lambda)$, which is similar to two-dimensional (2D) Dirac fermion in graphene~\cite{V.P.Gusynin}.  Since the energy spacing between the higher LLs is smaller, the associated $\sigma_{xy}$ will get suppressed at first, which will occur successively for the lower LLs.  In the case when the energy spacing is less than the strong $\eta$, $\sigma_{xy}$ becomes negative. 

\begin{figure*} 
	\centering
	\includegraphics[width=17.2cm]{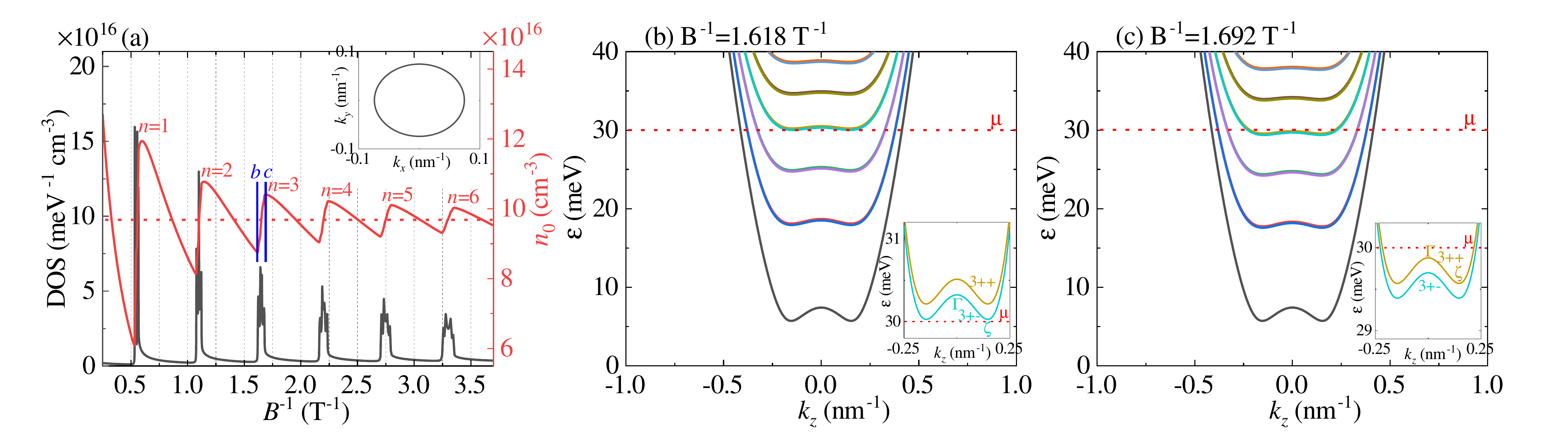}
	\caption{(Color online)  The DOS and carrier density $n_0$ versus the inverse magnetic field $B^{-1}$ with the fixed chemical potential $\mu=30$ meV.  When the magnetic field is absent, the red dotted line indicates the position of the carrier density, and the inset shows the extremal Fermi surface in the $k_x$-$k_y$ plane.  (b), (c) The LL dispersions versus $k_z$, with the magnitude of $B^{-1}$ corresponding to the blue lines $b$ and $c$ labeled in (a), respectively.  The insets are the enlarged plots of the dispersions around $\mu$.  We set $\eta=0.01$ meV. }
	\label{Fig3}
\end{figure*}

Then we study the resistivity.  To make direct comparisons with experiments, the resistivities are plotted as a function of the magnetic field $B$ in Figs.~\ref{Fig2}(c) and~\ref{Fig2}(d).  With weak $\eta=0.01$ meV, the conductivities satisfy $\sigma_{xx}\ll\sigma_{xy}$, and correspondingly, the longitudinal resistivity and Hall resistivity are given as 
\begin{align}
\rho_{xx}=\sigma_{xx}\sigma_{xy}^{-2}, \quad 
\rho_{xy}=\sigma_{xy}^{-1}, 
\end{align} 
respectively.  Thus in Fig.~\ref{Fig2}(c), $\rho_{xx}$ exhibits the SdH oscillations on an increasing background, which keeps the features of $\sigma_{xx}$.  The fourfold peak structure of $\rho_{xx}$ is more clearly seen in the inset of Fig.~\ref{Fig2}(c).  On the other hand, $\rho_{xy}$ shows a linear dependence on $B$.  The linear $\rho_{xy}$ was derived by Abrikosov for Dirac electrons occupying the lowest zeroth LL~\cite{Abrikosov}, where the large and nonsaturated magnetoresistivity was understood with fixed carrier density.  In experiment, such a linear $\rho_{xy}$ was observed in the Weyl semimetal NbP~\cite{C.Shekhar} and the Dirac semimetal Cd$_3$As$_2$~\cite{L.P.He, T.Liang2015} even at a very low magnetic field. 

At strong $\eta=5$ meV, $\sigma_{xx}$ becomes comparable to $\sigma_{xy}$.  When the chemical potential crosses the $n$th LLs, the fourfold peaks in $\sigma_{xx}$ merge into the single peak structure.  Now in Fig.~\ref{Fig2}(d), for $\rho_{xx}$, the SdH oscillations are blurred but the kinks are seen.  For $\rho_{xy}$, it exhibits a ramp in the quantum limit that gradually increases with $B$.  Note that $\rho_{xy}$ shows a plateau in a broad range $0.604$ T$<B<1.186$ T.  In fact, such a plateau originates from the plateaus in $\sigma_{xx}$ and $\sigma_{xy}$ that are induced by impurity scatterings, as indicated by the arrows in Figs.~\ref{Fig2}(a) and~\ref{Fig2}(b).  These results strongly suggest that 3D QHE cannot be explained with fixed carrier density.

\subsection{Fixed chemical potential} 

\begin{figure*}
	\centering
	\includegraphics[width=12cm]{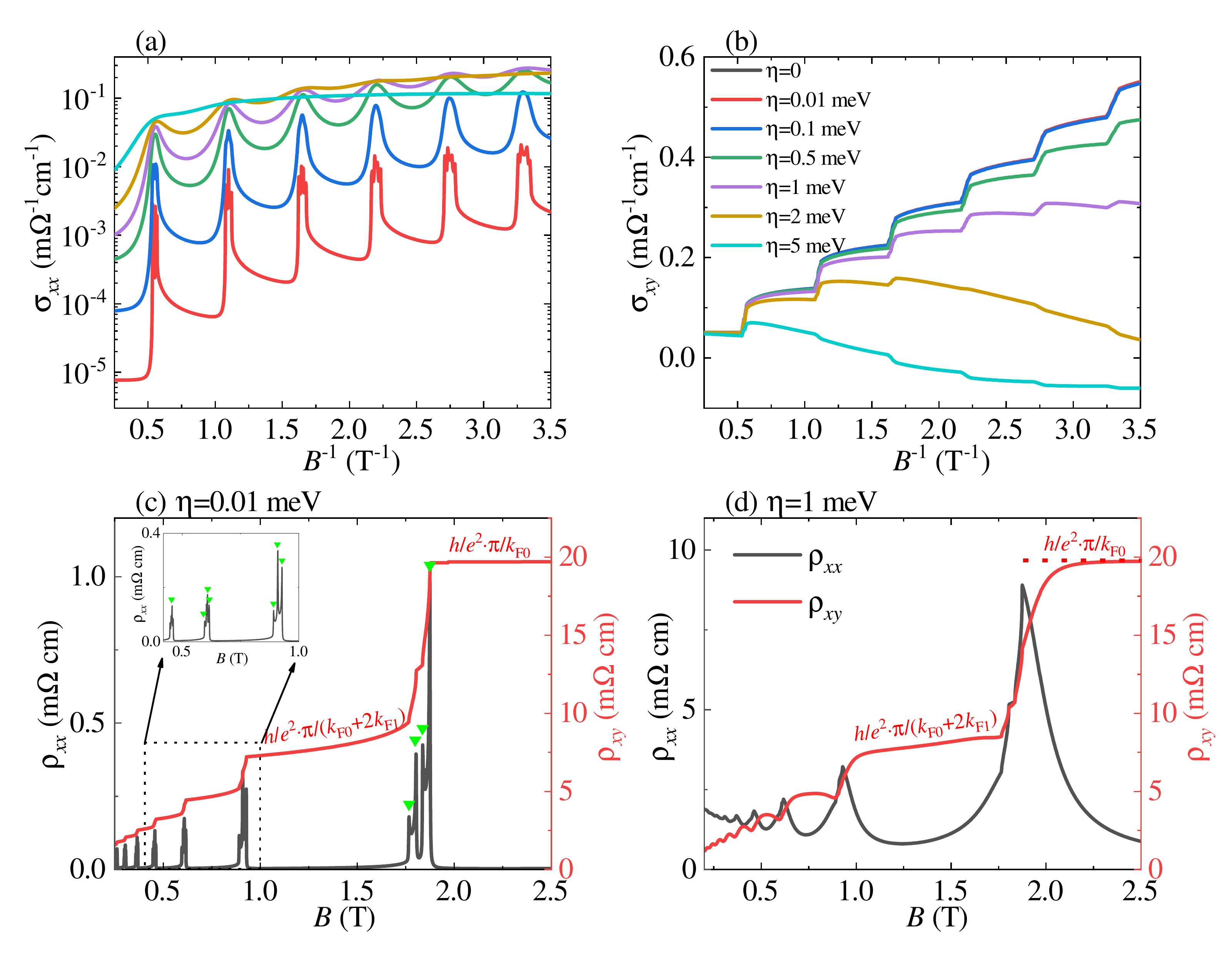}
	\caption{(Color online) The longitudinal conductivity $\sigma_{xx}$ and Hall conductivity $\sigma_{xy}$ versus the inverse magnetic field $B^{-1}$ for different linewidths $\eta$ in (a) and (b).  The longitudinal resistivity $\rho_{xx}$ and Hall resistivity $\rho_{xy}$ versus the magnetic field $B$ in (c) and (d), with the linewidth $\eta=0.01$ meV and $\eta=1$ meV, respectively.  In (c), the green triangles indicate the fourfold peak structure and the inset shows enlarged plot framed by the dotted line.  The legends are the same in (a) and (b), as well as in (c) and (d).  The fixed chemical potential is $\mu=30$ meV. }
	\label{Fig4}
\end{figure*}

In this section, we study the magnetotransport of ZrTe$_5$ under the condition of fixed chemical potential, which is set as $\mu=30$ meV.  

The calculated DOS and carrier density are plotted in Fig.~\ref{Fig3}(a).  The quantum limit is reached at the critical magnetic field $B_c=1.88$ T.  Now as the $\Gamma$ point of the zeroth LL is always below the chemical potential, 
\begin{align}
\varepsilon_{0+}(\Gamma)=M-\frac{\xi}{l_B^2}<\mu, 
\end{align}
the anomalous peak that appears in Fig.~\ref{Fig1}(a) is absent.  In the QO regime, we can see that the fourfold peaks of the DOS are overlapping and thus are difficult to be distinguished.  On the other hand, the carrier density oscillates with a period $\Delta_{1/B}=0.55$ T, implying that the  charging energies exist in the system.  The period can also be understood with the Onsager's relation.  Without a magnetic field, the extremal Fermi surface in the $k_x$-$k_y$ plane is plotted in the inset of Fig.~\ref{Fig3}(a), with its area $S_e=0.01724$ nm$^{-2}$.  Then, according to Eq.~(\ref{period}), we have $\Delta_{1/B}^O=0.556$ T$^{-1}$, which shows good consistency. 

The competition between the local DOS and the uniform DOS that was used to analyze the chemical potential oscillations in the previous section can be extended to illustrate the carrier density oscillations here.  It is worth pointing out that the oscillations in Figs.~\ref{Fig1}(a) and~\ref{Fig3}(a) are out of phase.  For example, in the insets of Figs.~\ref{Fig1}(b) and~\ref{Fig3}(b), although both the chemical potentials $\mu$ meet the $\zeta$ point of the $(3+-)$ LL, in Fig.~\ref{Fig1}(a), $\mu$ behaves as a peak, whereas in Fig.~\ref{Fig3}(a), the corresponding carrier density behaves as a valley.  

Next we turn to the conductivity and resistivity.  Figures~\ref{Fig4}(a) and~\ref{Fig4}(b) show that the effects of the increasing impurity scatterings $\eta$ on the conductivities $\sigma_{xx}$ and $\sigma_{xy}$ are the same as those in Figs.~\ref{Fig2}(a) and~\ref{Fig2}(b), respectively.  Figure~\ref{Fig4}(c) shows that with weak linewidth $\eta=0.01$ meV, in the longitudinal resistivity $\rho_{xx}$, the SdH oscillations are evident, but the fourfold peaks can only be distinguished for the $n=1$ LLs and are overlapping for other LLs, with  details seen in the inset of Fig.~\ref{Fig4}(c).  This means that the saddle points of the inverted LLs cannot be well distinguished with fixed chemical potential~\cite{L.You}, even when the spin Zeeman effect is included.  In the Hall resistivity $\rho_{xy}$, the quasi-quantized plateaus are clearly observable. 

Actually, a 3D Hall system under a magnetic field can be regarded as the 2D slice stacking~\cite{A.A.Burkov2011, Y.X.Wang2019}.  Then the Hall conductivity (in unit of $\sigma_0$) is a summation of the Chern numbers of the occupied LLs (see Appendix C), 
\begin{align}
\sigma_{xy}=\sigma_0\sum_{ns\lambda}\int_{-\pi}^\pi \frac{dk_z}{2\pi} 
C_{ns\lambda}^{k_z} \theta[\varepsilon_{ns\lambda}(k_z)<\mu], 
\label{sigmaxy-Chern}
\end{align}
in which the Chern number of the $ns\lambda$ LL at wave vector $k_z$ is $C_{ns\lambda}^{k_z}=1$ for Dirac fermions.  In the quantum limit, the Fermi wave vector $k_{F0}$ is obtained as 
\begin{align}
k_{F0}\simeq\sqrt{\frac{\mu+M}{\xi_z}-\frac{\xi}{\xi_zl_B^2}}
\simeq\sqrt{\frac{\mu+M}{\xi_z}}
=0.43 \text{ nm}^{-1}.
\label{kF0}
\end{align}
in which the $B$-dependent term $\frac{\xi}{\xi_zl_B^2}$ is much weaker than $\frac{\mu+M}{\xi_z}$ and can be neglected.  Then we have $\sigma_{xy}=\sigma_0\frac{k_{F0}}{\pi}$, leading to the quasi-quantized plateau in $\rho_{xy}$ that is estimated to be $\rho_{xy}=\sigma_{xy}^{-1}=18.9$ m$\Omega$ cm, as shown in Fig.~\ref{Fig4}(c).  For the $n\geq1$ LL, if occupied, the Fermi wavevector $k_{Fn}$ is obtained as
\begin{align}
k_{Fn}\simeq\Big(\frac{1}{\xi_z}\sqrt{\mu^2-\frac{2nv^2}{l_B^2}}+\frac{M}{\xi_z} \Big)^\frac{1}{2}. 
\label{kFn}
\end{align}
Here, as the $B$-dependent term cannot be neglected, $k_{Fn}$ decreases with $B$.  Thus in Fig.~\ref{Fig4}(c), the $n\geq1$ plateau of $\rho_{xy}$, which is given as $\rho_{xy}=\pi\sigma_0^{-1}(k_{F0}+2\sum_n k_{Fn})^{-1}$, with the factor $2$ accounting for the two LL branches, increases slowly with $B$.  For the kink of $\rho_{xy}$ around $B=1.8$ T in Figs.~\ref{Fig4}(c) and~\ref{Fig4}(d), it occurs when the $\lambda=-1$ branch is occupied and the $\lambda=1$ branch is unoccupied, with the magnitude $\rho_{xy}=\pi\sigma_0^{-1}(k_{F0}+k_{F1})^{-1}$.

In Fig.~\ref{Fig4}(d) when $\eta$ increases to 1 meV, we observe that in $\rho_{xx}$, the fourfold peaks no longer exist, but the SdH oscillations are still discernible, while in $\rho_{xy}$, the Hall plateaus are preserved, with the neighboring plateau transitions becoming smooth.  More importantly, the nonvanishing dips in $\rho_{xx}$ correspond to the plateaus in $\rho_{xy}$, and the peaks in $\rho_{xx}$ correspond to the plateau transition regions in $\rho_{xy}$.  Although the peak heights in $\rho_{xx}$ and the plateau magnitudes in $\rho_{xy}$ depend heavily on the model parameters, the numerical results are qualitatively consistent with the behaviors of $\rho_{xx}$ and $\rho_{xy}$ observed experimentally in ZrTe$_5$~\cite{F.Tang} and HfTe$_5$~\cite{P.Wang}, which thus capture the main features of 3D QHE. 

In a 2D electron system, the magnetotransport is often described by fixed chemical potential, resulting in the well-defined Hall plateaus~\cite{V.P.Gusynin, D.N.Sheng}.  Here also under the condition of fixed chemical potential, we can understand 3D QHE as a close analog to 2D QHE, in which the quasi-quantized plateaus in $\rho_{xy}$ are  determined by the dependence of the Fermi wavevectors on the magnetic field.  Moreover, the impurity scatterings play indispensable roles in driving 3D QHE, since they can enhance the peak heights in $\rho_{xx}$ as well as smoothen the plateau transitions in $\rho_{xy}$.  Although our results based on a non-interacting model are quite different from the interaction-induced CDW scenario~\cite{F.Tang, P.Wang, F.Qin}, but are supported by a recent study~\cite{S.Galeski2021}, in which the CDW states are demonstrated to be absent in ZrTe$_5$. 

Note that in a Dirac semimetal $(\xi=\xi_z=0)$ with fixed chemical potential, the Fermi wavevectors are solved as 
\begin{align}
k_{F0}=\frac{1}{v_z}\sqrt{\mu^2-M^2}, 
\label{kF02}
\end{align}
for the $n=0$ LL, and
\begin{align}
k_{Fn}=\frac{1}{v_z}\sqrt{\mu^2-M^2-\frac{2nv^2}{l_B^2}},
\label{kFn2}
\end{align}
for the $n\geq1$ LL.  We see that the Fermi wavevectors in Eqs.~(\ref{kF02}) and~(\ref{kFn2}) show similar dependence on the magnetic field as those in Eqs.~(\ref{kF0}) and~(\ref{kFn}).  Based on these results, we  suggest that the proposed mechanism for 3D QHE is justified for a 3D Dirac fermion system, no matter the ground state is a strong TI ($\xi>0$ and $\xi_z>0$), a weak TI ($\xi>0$ and $\xi_z<0$), or even a Dirac semimetal.

\section{Conclusions}

To summarize, we have numerically studied QOs and 3D QHE in ZrTe$_5$ by using the strong TI model. Under the conditions of fixed carrier density and fixed chemical potential, although both QOs can be explained in a similar way and the impurity scatterings show a similar effects on the conductivities, the main conclusions are totally different.  

With fixed carrier density, we find that there exists a deferring effect in the chemical potential, thus the saddle points of the inverted LLs are well distinguished by tuning the magnetic field magnitude.  This can provide an effective route beyond the surface states to identify the nontrivial topological phase.  
On the other hand, with fixed chemical potential, we demonstrate that 3D QHE originates from the interplay between Dirac fermions, magnetic field and impurity scatterings, which may represent a universal mechanism of the 3D Dirac fermion systems and needs more studies in the future.

\section{Acknowledgment} 

This work was supported by the Natural Science Foundation of China (Grant No. 11804122), the China Postdoctoral Science Foundation (Grant No. 2021M690970).

\section{Appendix}

\subsection{Derivation of Eqs.~(\ref{sigmaxx}) and~(\ref{sigmaxy}) in the main text}

The conductivity tensors can be derived from the Kubo-Streda formula in Eq.~(\ref{Kubo-Streda}).  
At zero temperature, the longitudinal conductivity $\sigma_{xx}$ is written as 
\begin{align*}
\sigma_{xx}&=-\frac{1}{\pi V}
\sum_{\boldsymbol k}\int d\varepsilon \frac{df(\varepsilon)}{d\varepsilon} 
\text{Tr} \Big[J_x \text{Im}G^R(\varepsilon) J_x \text{Im}G^R(\varepsilon) \Big]
\\
&=\frac{1}{\pi V}\sum_{\boldsymbol k}
\text{Tr} \Big[J_x \text{Im}G^R(\mu) J_x \text{Im}G^R(\mu) \Big], 
\tag{A1}
\label{sigmaxxA1}
\end{align*}
in which the identity $\frac{df(\varepsilon)}{d\varepsilon}=-\delta(\varepsilon-\mu)$ is used in the second line.  Note that the expression of $\sigma_{xx}$ in Eq.~(\ref{sigmaxxA1}) is the same as those used in previous studies~\cite{H.W.Wang, Y.X.Wang2020}.  Consider the LL degeneracy $g=\frac{1}{2\pi l_B^2}$, and insert the energies and wavefunctions of the LLs, then we have 
\begin{align*}
\sigma_{xx}
&=\frac{g\eta^2}{\pi L_z}\sum_{k_z}\sum_{ns\lambda}\sum_{n's'\lambda'} 
|\langle ns\lambda|J_x|n's'\lambda'\rangle|^2
\nonumber
\\
&\times[(\mu-\varepsilon_{ns\lambda})^2+\eta^2]^{-1}
[(\mu-\varepsilon_{n's'\lambda'})^2+\eta^2]^{-1}.  
\tag{A2}
\label{sigmaxxA2}
\end{align*}

\begin{figure*}
	\includegraphics[width=12.2cm]{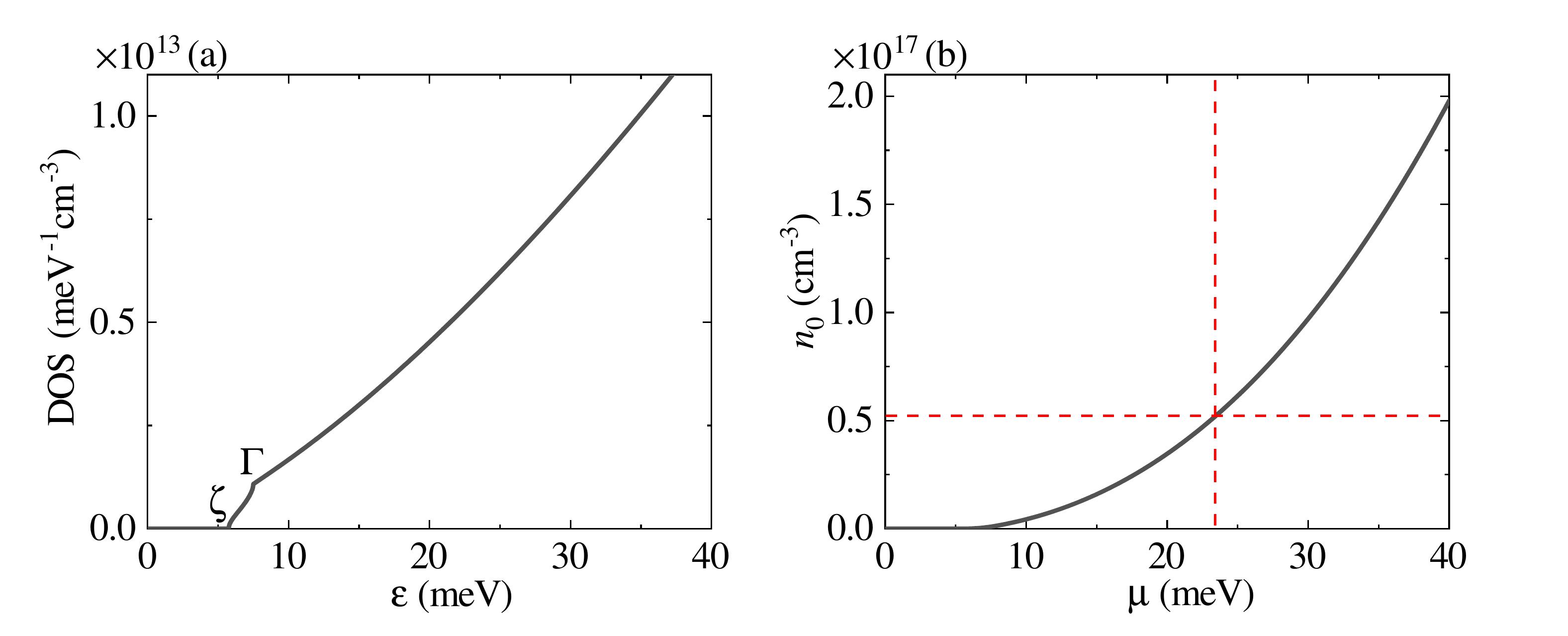}	
	\caption{(Color online) (a) The DOS versus the energy $\varepsilon$, and (b) the carrier density $n_0$ vs the chemical potential $\mu$.  The linewidth broadening is set as $\eta=0.01$ meV.  The kinks in (a) are related to the saddle points, the $\Gamma$ point and the $\zeta$ point.  The red dashed lines in (b) show that $n_0=5.2\times10^{16}$ cm$^{-3}$ corresponds to $\mu_0=23.41$ meV.}
	\label{FigA1}	
\end{figure*} 

Similarly, the Hall conductivity $\sigma_{xy}$ is 
\begin{widetext}
\begin{align*}
\sigma_{xy}
=&\frac{g}{2\pi L_z}\sum_{k_z}\sum_{ns\lambda}\sum_{n's'\lambda'} 
\int d\varepsilon f(\varepsilon) \Big[\langle ns\lambda| 
J_x\frac{dG^R}{d\varepsilon}|n's'\lambda'\rangle \langle n's'\lambda'|J_y (G^A-G^R)|ns\lambda\rangle 
-\langle ns\lambda|J_x (G^A-G^R)|n's'\lambda'\rangle
\\
&\times \langle n's'\lambda'|J_y\frac{dG^A}{d\varepsilon}|ns\lambda\rangle \Big]
\\
=&\frac{g}{L_z} \sum_{k_z}\sum_{ns\lambda}\sum_{n's'\lambda'} 
\text{Im}\Big[\frac{\langle ns\lambda|J_x|n's'\lambda'\rangle 
\langle n's'\lambda'|J_y|ns\lambda\rangle}
{(\varepsilon_{ns\lambda}-\varepsilon_{n's'\lambda'}+i\eta)^2}
-\frac{\langle ns\lambda|J_y|n's'\lambda'\rangle \langle n's'\lambda'|J_x|ns\lambda\rangle }
{(\varepsilon_{ns\lambda}-\varepsilon_{n's'\lambda'}-i\eta)^2}\Big]
\theta(\mu-\varepsilon_{ns\lambda})
\theta(\varepsilon_{n's'\lambda'}-\mu). 
\tag{A3}
\label{sigmaxyA3}
\end{align*}
\end{widetext}
In previous literatures, the expressions of $\sigma_{xy}$ are much more complicated~\cite{H.W.Wang, B.Fu, C.Wang} and include a normal part and an anomalous part, where the former is determined by the states around the Fermi level and the latter reflects the thermodynamic property of the system.  Here Eq.~(\ref{sigmaxyA3}) is quite concise and thus are facilitate to the numerical calculations.   

After calculating the current density matrix elements $\langle ns\lambda|J_{x/y}|n's'\lambda'\rangle$, we obtain 
\begin{widetext}
	\begin{align}
	\sigma_{xx}=&\sigma_0\frac{2g\eta^2}{L_z} 
	\sum'\frac{M_{ns\lambda;n+1,s'\lambda'}^2}
	{[(\mu-\varepsilon_{ns\lambda})^2+\eta^2][(\mu-\varepsilon_{n+1,s'\lambda'})^2+\eta^2]}
	+\sigma_0\frac{2g\eta^2}{L_z} 
	\sum''\frac{M_{ns\lambda;n-1,s'\lambda'}^2}
	{[(\mu-\varepsilon_{ns\lambda})^2+\eta^2][(\mu-\varepsilon_{n-1,s'\lambda'})^2+\eta^2]}, 
    \tag{A4}	
	\label{sigmaxxA4}
\end{align}	
and 
	\begin{align}
	\sigma_{xy}=&\sigma_0\frac{4g\pi}{L_z} \sum'
	\frac{(\varepsilon_{ns\lambda}-\varepsilon_{n+1,s'\lambda'})^2-\eta^2}
	{[(\varepsilon_{ns\lambda}-\varepsilon_{n+1,s'\lambda'})^2+\eta^2]^2}
	M_{ns\lambda;n+1,s'\lambda'}^2
	\theta(\mu-\varepsilon_{ns\lambda}) 
	\theta(\varepsilon_{n+1,s'\lambda'}-\mu)
	\nonumber\\
	&-\sigma_0\frac{4g\pi}{L_z} \sum''
	\frac{(\varepsilon_{ns\lambda}-\varepsilon_{n-1,s'\lambda'})^2-\eta^2}
	{[(\varepsilon_{ns\lambda}-\varepsilon_{n-1,s'\lambda'})^2+\eta^2]^2}
	M_{ns\lambda;n-1,s'\lambda'}^2
	\theta(\mu-\varepsilon_{ns\lambda}) 
	\theta(\varepsilon_{n-1,s'\lambda'}-\mu),
    \tag{A5}	 
	\label{sigmaxyA5}
	\end{align}  
\end{widetext}
where the summation signs are 
\begin{align}
\sum'=\sum_{k_z}\sum_{n\geq0,s\lambda} \sum_{s'\lambda'},  
\quad
\sum''=\sum_{k_z}\sum_{n\geq1,s\lambda} \sum_{s'\lambda'}, 
\tag{A6}
\end{align} 
and the matrix element $M_{ns\lambda;n+1,s'\lambda'}$ is given in Eq.~(\ref{matrixelement}). 

We note that the components of the conductivity satisfy the following relations: 
\begin{align}
&\sigma_{xx}(ns\lambda\rightarrow n+1,s'\lambda')
=\sigma_{xx}(n+1,s'\lambda'\rightarrow ns\lambda),
\tag{A7}
\\
&\sigma_{xy}(ns\lambda\rightarrow n+1,s'\lambda')
=-\sigma_{xy}(n+1,s'\lambda'\rightarrow ns\lambda),
\tag{A8}
\end{align} 
implying that the contributions to $\sigma_{xx}$ from the LL transition $ns\lambda\rightarrow (n+1,s'\lambda')$ and from $(n+1,s'\lambda')\rightarrow ns\lambda$ are equal, while those to $\sigma_{xy}$ are opposite.
According to these properties, we can change the index $n\rightarrow n+1$, $s\leftrightarrow s'$, and $\lambda\leftrightarrow\lambda'$ in the second summation of Eqs.~(\ref{sigmaxxA4}) and~(\ref{sigmaxyA5}), and obtain the final expressions in Eqs.~(\ref{sigmaxx}) and~(\ref{sigmaxy}) in the main text.

\subsection{Density of states and chemical potential with no magnetic field}

When the magnetic field is absent, the energies for the Hamiltonian $H(\boldsymbol k)$ in Eq.~(\ref{H0}) can be obtained directly, which are given as 
\begin{align}
\varepsilon_{s\boldsymbol k}=s\varepsilon_{\boldsymbol k}
=&s\Big[v^2 (k_x^2+k_y^2)+v_z^2 k_z^2+\big(M-\xi(k_x^2+k_y^2)
\nonumber\\
&-\xi_zk_z^2\big)^2\Big]^\frac{1}{2},
\tag{A9}
\label{bands}
\end{align}
where $s=\pm1$ denotes the conduction/valence band.  Note that both the conduction and valence bands own twofold degeneracy.  

The DOS is calculated as, 
\begin{align*}
D(\varepsilon>0)=\frac{g_b\eta}{8\pi^4} \iiint dk_x dk_y dk_z 
\frac{1}{(\varepsilon-\varepsilon_{\boldsymbol k})^2+\eta^2},  
\tag{A10}
\label{DOS1}
\end{align*}
where $g_b=2$ denotes the twofold degeneracy.  To do the integrations, we make the following substitutions~\cite{Y.X.Wang2020}: $x=vk_x$, $y=vk_y$, $z=v_zk_z$, and $\xi'=\frac{\xi}{v^2}$, $\xi_z'=\frac{\xi_z}{v_z^2}$, and change the Cartesian into the Cylindrical system: $x=\rho\text{cos}\theta$, $y=\rho\text{sin}\theta$.  After completing the integration over the azimuth angle $\theta$, we have
\begin{align*}
&D(\varepsilon)
=\frac{g_b\eta}{4\pi^3 v^2 v_z} 
\int_{-\infty}^\infty dz \int_0^\infty d\rho\rho
\\
&\times\Big[\Big(\varepsilon-\sqrt{\rho^2+z^2+(M-\xi'\rho^2-\xi_z' z^2)^2}\Big)^2+\eta^2\Big]^{-1}. 
\tag{A11}
\label{DOS2}
\end{align*}

On the other hand, the carrier density is defined in Eq.~(\ref{n0}), which, at zero temperature, is written as 
\begin{align*}
n_0&=\int_0^{\mu} d\varepsilon D(\varepsilon). 
\tag{A12}
\label{carrier}
\end{align*}

Evidently, the integrands in Eqs.~(\ref{DOS2}) and~(\ref{carrier}) show complicated dependencies on $\rho$ and $z$, thus the integration cannot be solved analytically.  But we can resort to the numerical calculations.  In Figs.~\ref{FigA1}(a) and~\ref{FigA1}(b), we plot the calculated DOS as a function of the energy $\varepsilon$ and the carrier density $n_0$ as a function of the chemical potential $\mu$, respectively.  We see that the DOS increase with $\varepsilon$ and $n_0$ increases with $\mu$, both in a monotonous tendency.  In Fig.~\ref{FigA1}(b), the red dashed lines denote that the chosen carrier density $n_0=5.2\times10^{16}$ cm$^{-3}$ in Sec. IVA corresponds to the chemical potential $\mu=23.41$ meV.

\subsection{Derivation of Eq.~(\ref{sigmaxy-Chern}) in the main text}

Here we give a detailed derivation of Eq.~(\ref{sigmaxy-Chern}).  
The 2D Chern number of the $n$-th band is defined as an integration over the BZ, 
\begin{align}
C_n&=-\frac{1}{2\pi i}\int_{\text{BZ}} d^2k \Omega_n, 
\tag{A13}
\end{align}
where $\Omega_n(\boldsymbol k)=\nabla_{\boldsymbol k}\times A_n(\boldsymbol k)\big|_z$ is the Berry curvature and $A_n(\boldsymbol k)=\langle u_n|\nabla_{\boldsymbol k}|u_n\rangle$ is the Berry connection.  Then we have
\begin{align}
C_n&=-\frac{1}{\pi}\text{Im}\sum_{n'}
\int_{\text{BZ}} d^2k
\langle \frac{\partial u_n}{\partial k_x}|u_{n'}\rangle
\langle u_{n'}|\frac{\partial u_n}{\partial k_y} \rangle
\nonumber\\  
&=-\frac{1}{\pi}\text{Im}\sum_{n'}
\int_{\text{BZ}} d^2k
\frac{\langle u_n|v_x|u_{n'}\rangle
\langle u_{n'}|v_y|u_n\rangle} {(\varepsilon_n-\varepsilon_{n'})^2},  
\tag{A14}
\label{Chern}
\end{align}
where $\varepsilon_n$ and $u_n$ are the energy and wave function of the $n$-th band.  Note that in the second line of Eq.~(\ref{Chern}), the identity $\langle\frac{\partial u_n}{\partial k_\alpha} |u_{n'}\rangle=\frac{\langle u_n|v_\alpha|u_{n'}\rangle} {\varepsilon_n-\varepsilon_{n'}}$ is used.  Comparing Eq.~(\ref{Chern}) with Eq.~(\ref{sigmaxyA3}), we obtain Eq.~(\ref{sigmaxy-Chern}) in the main text.

\end{document}